# In situ annealing of superconducting MgB$_2$ films prepared by pulsed laser deposition

Yue Zhao, Mihail Ionescu, Alexey V. Pan, Shi Xue Dou
Institute for Superconducting and Electronic Materials,
University of Wollongong, NSW 2522, Australia
E. W. Collings
Dept. of Materials Science and Engineering,
Ohio State University, Columbus, OH, 43210-1179, USA
E-mail: yz70@uow.edu.au

**Abstract:**
The in situ annealing conditions of pulsed laser deposited MgB$_2$ films were studied. The precursor films were deposited at 250℃ from a stoichiometric MgB$_2$ target in a 120mTorr Ar atmosphere. The films were then *in situ* annealed at a temperature from 450℃ to 800℃ and an annealing time from 1 minute to 10 minutes. We found that the superconducting properties depend in a crucial way on the annealing conditions: temperature, heating rate and time. The best film with a thickness of ~600nm was obtained under the following annealing conditions: T$_{anneal}$=680-690℃, t$_{anneal}$=1 min, heating rate= 38℃/min. The T$_{c\ onset}$ of the film is 28K with a transition width of ~10K. The hysteresis loop of magnetic moment of the film indicates weak field dependence in high fields. Magneto-optical imaging of the film showed quite homogeneous magnetic flux penetration, indicating structural homogeneity. The films without annealing showed no superconductivity.

## 1. Introduction

The discovery of superconductivity in MgB$_2$ in 2001[1] has already attracted tremendous interest and numerous investigations on this material. MgB$_2$ has a critical temperature (T$_c$) of 39K in zero field, which is much higher than that of Nb$_3$Sn. MgB$_2$ films were successfully grown by Pulsed Laser Deposition (PLD) technique [2,3] shortly after the discovery of superconductivity in this material. The supercurrent can easily pass the grain boundaries of MgB$_2$, which makes it different from the high-temperature superconductors [4]. Since MgB$_2$ is much cheaper than HTS cuprates, it is very attractive to develop superconducting electronics or coated conductors from this material.

However, due to difficulties in fabricating MgB$_2$ films, there is not yet an optimal method to produce superconducting films for applications. Obstacles in the formation of MgB$_2$ films come mainly from two factors: the high volatility of magnesium (vapor pressure ~1x10$^{-6}$Torr at 250℃ and 8.3Torr at 727℃) and its high reactivity with oxygen. Ueda et. al. have written an extensive review on the methods of MgB$_2$ film preparation [5]. Generally, these methods are classified into three types, as grown, *in situ* annealing and *ex situ* annealing. To date, good T$_c$ has been achieved by *ex situ* annealing [2,3,6,7]. The shortcoming of the *ex situ* annealing method is the difficulty in realizing multi-layer structures, which are essential for making junctions or other superconducting electronic devices. The as-grown method is capable of producing multi-layer structures. As grown MgB2 films with an applicable high T$_c$ of 35K have been obtained by molecular beam epitaxy (MBE)[8]. Some other method, such as PLD [9] and magnetic sputtering [10], can also achieve as grown superconducting MgB$_2$ films, but the T$_c$ properties are not as good as MBE. *In situ* high temperature annealing carried out immediately after the deposition has proven to be effective to obtain high T$_c$ [7,11-16]. The advantage of the *in situ* annealing method is that it can achieve multi-layer structures and high T$_c$ in the films simultaneously. Recently, Xi and co-workers obtained epitaxial MgB$_2$ films by hybrid physical-chemical vapor deposition (HPCVD) [17]. The T$_c$ value of their films reached the same level as bulk MgB$_2$. This is a significant improvement of the preparation of MgB$_2$ films. Further work on optimization of this as grown method may lead to superconducting electronic devices manufactured from this material.

As shown in the literature on *in situ* annealing of MgB$_2$ films, the annealing parameters are hard to select. The magnesium vapor pressure inside the deposition chamber cannot be made high enough to prevent the evaporation of magnesium. So, in order to keep the stoichiometry in the MgB$_2$ film, the annealing temperature should be low and the duration should be short. On the other hand, a reasonably high annealing temperature and long time is necessary for MgB$_2$ phase formation and crystallization in the films. Since the process behind the formation of the superconducting phase in the MgB$_2$ film still remains unclear, it is very interesting to study how the *in situ* annealing conditions influence the microstructure and superconducting properties of MgB$_2$ films.

In this paper, we applied different temperatures, heat treatment durations and heating rates on *in situ* annealing of PLD MgB$_2$ films. Optimal annealing conditions were obtained according to the T$_c$ measured in the prepared films. The annealing process is discussed with the help of the T$_c$ curves, XRD results, atomic force microscopy (AFM), magneto-optical images (MOI), and magnetization hysteresis curves.

## 2. Experimental details

The PLD process was conducted in a spherical chamber with a volume of ~52L. The stoichiometric MgB$_2$ target (84% density) and magnesium target were set on a carrousel in the chamber. We used sapphire-R (1$\bar{1}$02) substrates with dimensions of 3x3 mm$^2$, which is non reactive with the MgB$_2$ film at high temperature [18]. The substrates were mounted onto the heater with silver paste.

The laser beam was generated by an excimer laser system (Lambda-Physik) operating on KrF gas ($\lambda$=248nm, 25ns). The laser beam was focused to an elliptical spot of ~7x1.5mm by cylindrical lenses of 700mm focal length. The laser energy was 500-510mJ/pulse. Thus the fluence was ~6 J/cm$^2$. The chamber was evacuated to a base pressure of ~4x10$^{-7}$ Torr and then filled with high purity argon to 120mTorr as background gas. The taget-substrate distance is 40mm. During the deposition, the pulse frequency was 10Hz, and the heater was kept at 250℃.

Two groups of precursor films were made: the thicker group had a longer deposition time (15min) and a thickness of ~1μm. The thinner group had a shorter deposition time (5min) and a thickness of ~0.5μm. After the MgB$_2$ deposition, an Mg cap layer was deposited onto the film surface.

The *in situ* annealing process was conducted in a 760Torr argon atmosphere. The heater was heated to 450-800℃ and maintained at that temperature for 1~10min. Then the power was switched off, and the samples were cooled down at a rate of ~55℃/min when T>400℃. Two types of heating procedure were used, a rapid one to heat the sample at a rate of about 110℃/min ($t_{ramp}$=4min), and a slower one to heat the sample more slowly at various rates from 25℃/min to 63℃/min ($t_{ramp}$=7-17min). It was found difficult to keep the annealing temperature stable within 1min and there is usually an overshoot. We take the actual peak temperature as the annealing temperature.

To test the possibility of improving the quality of our $MgB_2$ film, we prepared an $MgB_2$ film with lower laser energy of 400-420mJ/pulse (~5 J/cm$^2$). In order to get a similar growth rate, the target-substrate distance was reduced to 30mm, and a thickness of ~600nm was obtained by a 5min $MgB_2$ deposition. The film was *in situ* annealed using the optimum conditions obtained from the annealing experiments.

$T_c$ of the films was obtained by DC magnetization measurement on a SQUID magnetometer system (MPMS, Quantum Design). The applied field was perpendicular to the film surface. We chose the $T_{c\ onset}$ as the point where the magnetic moment begins to drop, and $T_{c\ offset}$ as the point where the magnetic moment becomes constant. The magnetization loops up to 5T at 5K, 10K, 15K, 20K were also obtained in the same magnetometer with the magnetic field perpendicular to the film surface. XRD of the films was conducted on a M03XHF22 diffractometer (MAC Science). The surface microstructure and the thickness of the films were observed with AFM (SPM, Digital Instruments) using contact mode. The magneto-optical imaging was carried out in a home-made optical cryo-system. An epitaxial ferrite-garnet magneto-optical indicator film with in-plane magnetization was used for imaging. Energy-dispersive spectrometry (EDS, Oxford) with ultra-thin window was utilized to reveal the peaks of light elements in the films.

## 3. Results

$T_c$ measurements show that there is a quite narrow temperature window for annealing $MgB_2$ films (Fig. 1). The films shown in Fig. 1 were rapidly heated (110℃/min) during annealing. It is quite clear that the thicker films (~1μm) have higher $T_c$ than the thinner films(~0.5μm). The $T_{c\ onset}$ increased when the annealing temperature rose from ~640℃ to ~700℃, whereas the films became non-superconducting if annealing temperature was raised further. The highest $T_{c\ onset}$ for the thicker films is 29.6K (annealed at 720℃, 1min), and the highest $T_{c\ onset}$ for the thinner film is 27K (annealed at 690℃, 1min).

We examined the influence of the heating rate on $T_c$ of the thinner films at an annealing temperature between 680~690℃. As shown in Fig. 2, the $T_{c\ onset}$ of the film with slow heating rate (63℃/min) is 17K, rather lower than that of the rapid-heated (110℃/min) one. At a heating rate of 38℃/min, the $T_{c-onset}$ increases to 27K, and the $T_{c\ offset}$ increases to 8K. By further decreasing the heating rate to 25℃/min, the $T_c$ of the produced film drops to 20K. A reasonably slow heating rate of about 38℃/min seems to be beneficial to obtain high $T_{c\ onset}$ and a narrower transition width. A difficulty of $T_c$ comparison between rapid heated annealing and slow heated annealing is that the temperature lag between the thermal couple and the sample may change with different heating rate. The temperature calibration reveals that the thermal couple position is changing its temperature slower than the sample position. At very high heating rate, the sample temperature could be about 20  higher than the detected value from the thermal couple, whereas the this difference is much smaller (<5℃) at lower ramp rates such as 38℃/min. So the reason for the drop of $T_{c\,onset}$ from heating rate of 110℃/min to 63℃/min may be that the rapid heated film actually undergone higher peak temperature.

A longer annealing time, 10min at 640℃ and 800℃ with rapid heating was also attempted. However, the films were not superconducting. A probably reason could be that the magnesium cap layer is too thin for that circumstance. The 3min magnesium deposition provides a cap layer of 800nm thickness. We also carried out 15min depositions to put thicker cap layers, but still could not get superconducting film by 10 min annealing.

Based on the results described above, we consider that the optimal annealing conditions for our ~0.5μm $MgB_2$ film are to heat the films at a rate of ~38℃/min to 680~690℃ and hold the temperature for 1min, followed by free cooling down to room temperature.

We noticed that the pressure of the argon atmosphere during annealing could also be crucial to the superconducting properties of the $MgB_2$ films. A 760Torr background gas pressure is necessary to obtain superconducting films, and the films annealed with a low Ar pressure of ~120mTorr were non-superconducting. This may be because that the stronger scattering effect of the high background gas pressure is beneficial to keep a high local Mg vapor pressure near the surface of the $MgB_2$ film.

The film deposited at a lower energy (400~420mJ/Pulse) has a similar $T_{c\,onset}$ value (28 K) and a narrower transition width ($\Delta T$~10 K). Fig. 3 shows a comparison of the DC moment curves of three typical films measured in the zero-field cooled state. The three samples, a, b, and c are: a) 1μm film annealed at 720℃ for 1min, b) 0.5μm film annealed at 680℃ for 1min, and c) film deposited at lower energy and annealed at 680℃ (600nm). It can be seen that film a) has a slightly higher $T_{c\,onset}$, but the transition zone is wide, film b) has narrower transition width, and film c) has the narrowest transition width.

Fig. 4 shows the magneto-optical image of film c at 20 K and 8.7 mT. The regular rooftop pattern of the magnetic flux distribution reveals a homogeneous current flow all over the film.

The magnetization differences ($\Delta M$) in hysteresis loop of film c at 5K, 10K,15K and 20K are shown in Fig. 5. The applied magnetic field is perpendicular to the film surface. Considering that the magnetic flux penetrates in the film homogeneously in magneto optical imaging, standard expression of Bean model could be valid to obtain the critical current. $J_c$ values were calculated from the $\Delta M$-$B_a$ curve using the simple expression [19], $J_c(A/cm^2)=30\Delta M(emu/cm^3)/a(cm)$, where 2a=0.3cm is the dimension of the square film, $\Delta M=\Delta m/V$ is the magnetization difference, $\Delta m$ (emu) is the magnetic moment difference in the hysteretic loop, and $V=(0.3cm)^2 \times (6 \times 10^{-5}cm)$ is the volume of the film. Although the model neglects the $J_c$ difference within the sample and the $J_c$ value may be overestimated from the equation used, the result could provide an approximate $J_c$ value of the thin film [2]. The $J_c$ values are shown in the right axis of Fig. 5. The $J_c$-$B_a$ curve at 5K exhibits a flux jump region for fields lower than 0.2 T. At higher field, $J_c$ value shows a weak field dependence and remain about $10^5 A/cm^2$ at 5 T. The $J_c$ value at 20K has a high value at zero field, but drops sharply with an increase in the applied field. However, at $B_a \geq 0.2$ T, the $J_c$ decreases much more slowly, indicating strong pinning at high fields.

The XRD patterns of all the *in situ* annealed films did not show obvious peaks other than those from the $Al_2O_3$-R substrate. This indicates that the *in situ* annealed $MgB_2$ films have very small grains. The lack of crystalline orientation in the thin film may also contribute to the absence of any MgB2 peaks [16].

The surface topography of the films was observed by AFM. Fig. 6(a) is the AFM image of film c. An average roughness of $R_a$~60nm was detected in the 5x5μm$^2$ range. The scan in Fig.6(b) shows more details of a island surface in 500x500nm$^2$ range. Contrary to the ex situ annealed MgB2 films, the surface topography does not show any crystallite with regular facets. The similar roughness and island feature were also observed in other our in situ annealed ~0.5 μm thick films. In order to apply the films to devices, significant improvement of surface smoothness is very necessary.

## 4. Discussion

Although the reason for the depressed $T_{c\ onset}$ and wide transition width of the films remains an open question, it seems that the depression is related to the annealing conditions. A common view on the depression of $T_c$ in $MgB_2$ films is that the magnesium is easily lost during deposition and *in situ* annealing. This could result in the formation of $MgB_4$, $MgB_6$, $MgB_7$, or $MgB_{12}$ in the film[10,20]. Hinks et. al. have demonstrated that $MgB_2$ is a line compound in the phase diagram instead of a solid solution [21]. As the result, the formation of $Mg_{1\pm x}B_2$ is unlikely to be an explanation for the decrease in $T_c$. According to Saito and co-workers, Mg-deficient $MgB_2$ films do not present a $T_c$ lower than that of the stoichiometic ones. The result is consistent with Hinks' work, which also shows that the $T_c$ value does not change significantly in Mg-deficient $MgB_2$ bulks [21]. So, further work still seems necessary to clarify the influence of Mg deficiency on the $T_c$ of $MgB_2$ films. In any case, if too much magnesium is lost in the films during deposition and annealing, it is hard to form the $MgB_2$ phase in the first place, and the film will become non-superconducting. That is probably why we obtain non-superconducting films at higher annealing temperature or longer annealing times. Poor crystallization may also affect the $T_c$ value of the $MgB_2$ film [9]. The relatively low temperature and short annealing time of *in situ* annealing seem likely to result in poor crystallization of $MgB_2$ films and make the films non superconducting.

Impurities, like oxygen and carbon (from pump oil) in the $MgB_2$ lattice or MgO phase precipitated from the matrix, may also contribute to the depression of the $T_c$ value [2]. In the EDS spectrum of the as grown MgB2 film, specially grown on silicon substrate for oxygen detection, a significant oxygen peak as well as a small carbon peak was revealed. On the other hand, the impurities could be beneficial for better vortex pinning [2,12,22,23]. The presence of impurities in the films may support the weak field dependence of $J_c$ in high fields.

The as-grown films in our experiments are not superconducting, but after a short post-deposition annealing at 640~730℃, the films became superconducting. This indicates that the formation of $MgB_2$ and some degree of crystallization occurs during the in situ annealing process.

Considering the extremely high temperature in the PLD plume, most of the species arriving at the substrate should be atoms or ions of magnesium and boron or B-rich compounds such as $MgB_4$ or $MgB_7$ instead of $MgB_2$ molecules. Due to the relatively low substrate temperature and high deposition rate, the mobility and diffusion of magnesium and boron atoms could be quite low, and the reaction forming $MgB_2$ was suppressed, as well as the crystallization. That is probably why no superconductivity was observed in the as grown films in our experiments.

When the annealing begins, the competition of two processes could be triggered. One process is the evaporation of magnesium in the film. The other process is the formation of $MgB_2$ phase. This competition is crucial for the selection of in situ annealing conditions. By keeping the magnesium in the films during annealing, the formation of $MgB_2$ phase is favored. From the $T_c$ change versus heating rate in Fig. 2, it appears that $MgB_2$ phase is better formed by keeping the $MgB_2$ film at a lower temperature for a longer time before heating it to a higher temperature.

To analyze the $MgB_2$ phase evolution in the film during annealing, it is necessary to refer to the $MgB_2$ phase diagram. The phase diagram of the magnesium-boron system has been studied by Liu et al using thermodynamic calculations [24]. Their results showed that the magnesium vapor pressure decides the equilibrium solid phase in the Mg-B binary system. At ~700℃ the $MgB_2$ phase can remain at a magnesium vapor pressure higher than $1 \times 10^{-2}$ Torr. If the magnesium vapor pressure is lower than this value, the equilibrium solid phase would change to $MgB_4$ or other compounds with a low magnesium content. This critical vapor pressure is, however, too high to realize inside a PLD chamber. So, there is always a risk of losing too much magnesium during the *in situ* annealing process at ~700 .

The magnesium cap layer can provide Mg vapor for a limited period at high temperature. According to observations, the 800nm-thick magnesium layer disappeared within <1min when heated above 500℃. The free magnesium in the film, which has not reacted with boron, could also be evaporated at the annealing temperature of about 700℃. After the sources of magnesium are exhausted, the $MgB_2$ phase formed in the film begins to decompose. In our experiment, the films annealed at 800℃ were not superconducting, indicating that the $MgB_2$ phase decomposed significantly within 1min at that temperature, even for the films with a thick magnesium cap layer. The exhausting of magnesium cap layer and decomposition of $MgB_2$ may also explain the $T_c$ drop at very slow heating rate during annealing.

At low annealing temperatures, the problem is likely to be the poor Mg-B reactivity and poor crystallization of the film. The diffusion coefficients of Mg and B atoms are comparatively low at lower temperatures. According to our experiments, annealing at 650℃ for 1min is not enough to form the $MgB_2$ phase. However, some groups obtained superconducting films by *in situ* annealing at 600℃ for just a few minutes or even seconds [13]. Probably the differences in microstructure and density of the precursor films resulted in the different annealing behavior.

## 5. Conclusion

We obtained superconducting $MgB_2$ films by *in situ* annealing stoichiometric PLD films with an Mg cap layer. The best $T_{c\,onset}$ for 1μm films is 29.6 K, and $T_c$ for 0.5μm films is 27K with a wide transition zone ($\Delta T>15K$). The presence of superconductivity in the films is very sensitive to the *in situ* annealing conditions. The annealing temperature range is rather narrow from 648℃ to 730℃. The annealing time should also be limited to several minutes. A moderate heating rate for annealing achieved higher $T_c$ and narrower

ΔT. The films deposited at 5J/cm$^2$ with our optimal *in situ* annealing conditions showed reasonably high $T_c$ values ($T_{c\ onset}$=28K, $\Delta T$~10K), weak $J_c$ field dependence and homogeneous superconducting properties.

**Acknowledgment**
This work is supported by Australian Research Council (ARC) under a Linkage Project (LP0219629) cooperating with Alphatech International and The Hyper Tech Research Inc.

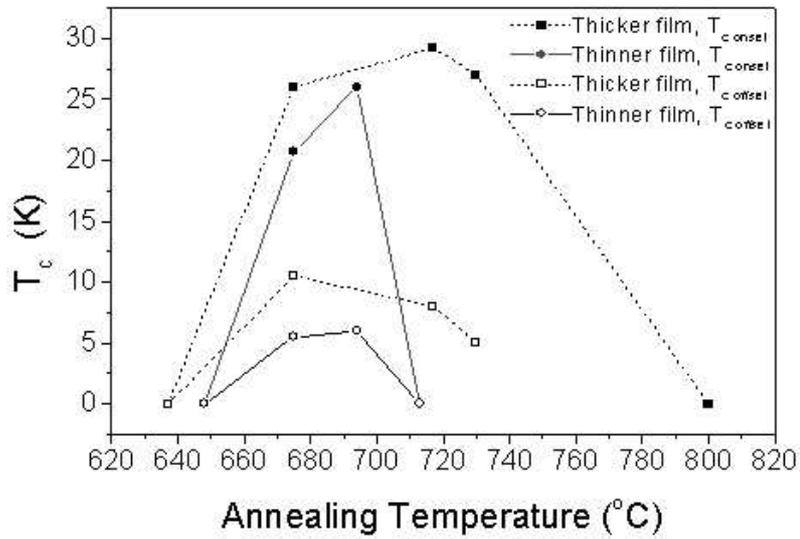

Fig. 1 $T_c$ values of $MgB_2$ films versus *in situ* annealing temperature. The thickness of the thicker films is ~2μm; The thickness of the thinner films is ~1μm. The heating rate of both group of films was ~110℃/min (rapid heating) with a 1min holding time.

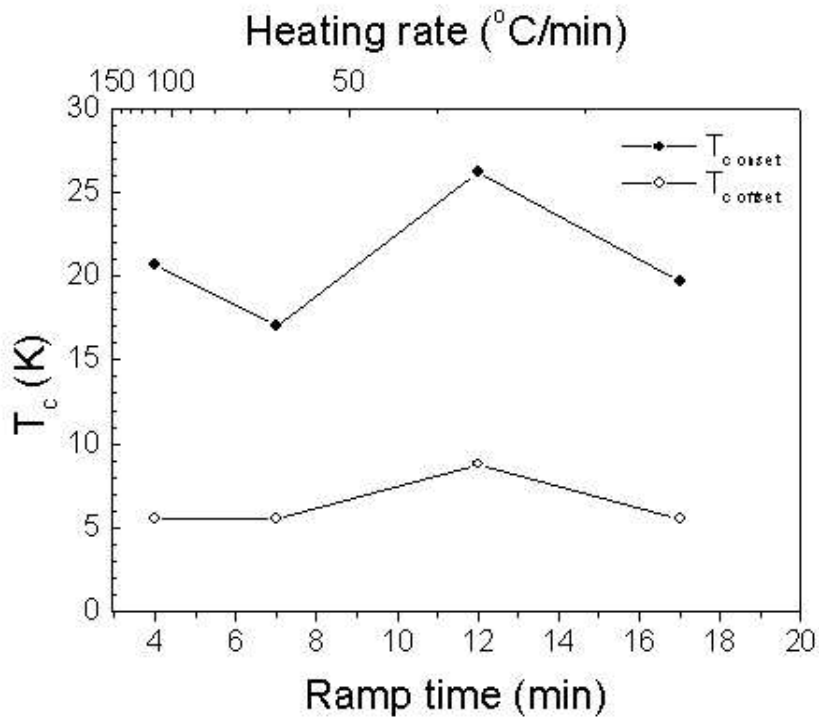

Fig. 2 $T_c$ values of $MgB_2$ films versus *in situ* annealing ramp time(bottom axis) and heating rate(top axis). The annealing temperature is ~680℃ with 1min holding time.

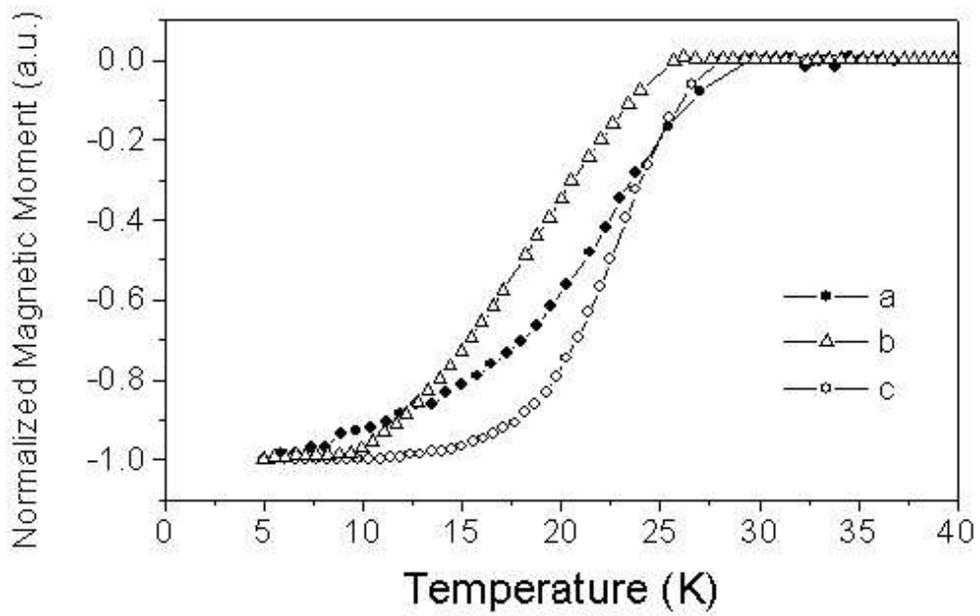

Fig.3. Magnetization curves of three typical $MgB_2$ films, samples a, b and c: a) $6J/cm^2$ deposition, 2μm thickness, 110℃/min heating, 720℃ 1min annealing; b) $6J/cm^2$ deposition, 38℃/min heating, 1μm thickness, 680℃ 1min annealing; c) $5J/cm^2$ deposition, 600nm thickness, 38℃/min heating, 680℃ 1min annealing.

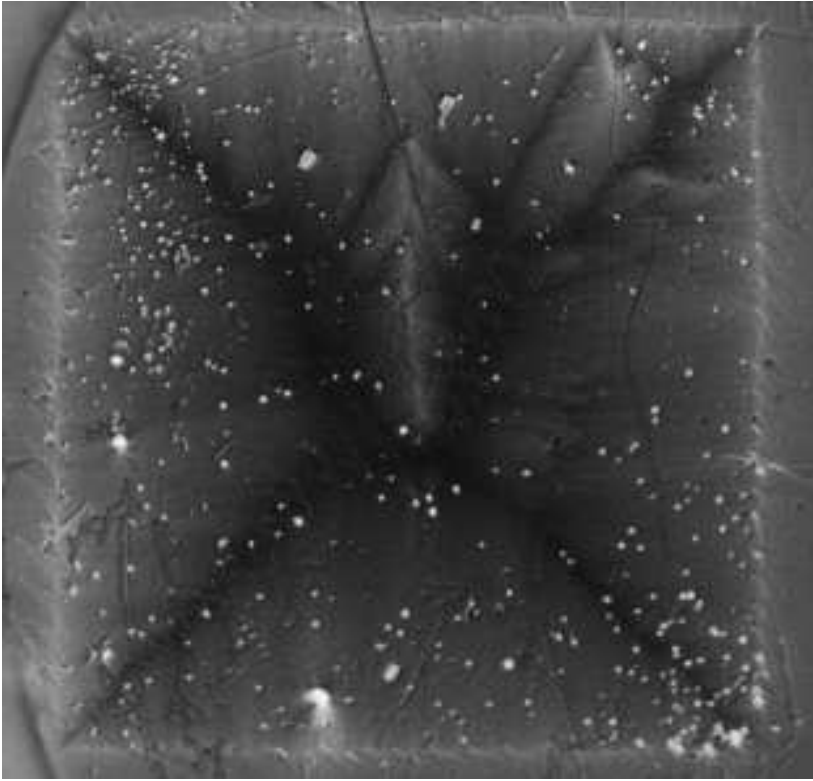

Fig.4. Magneto-optical image of film c at $B_a$=8.7mT at 20K after zero field cooling. The film size is 3x3mm². The brighter the image, the larger the magnetic flux. The white round spots are defects in the MO indicator. The bright area in the upper middle part of the film is an enhanced flux penetration due to an accidental mechanical scratch on the $MgB_2$ film.

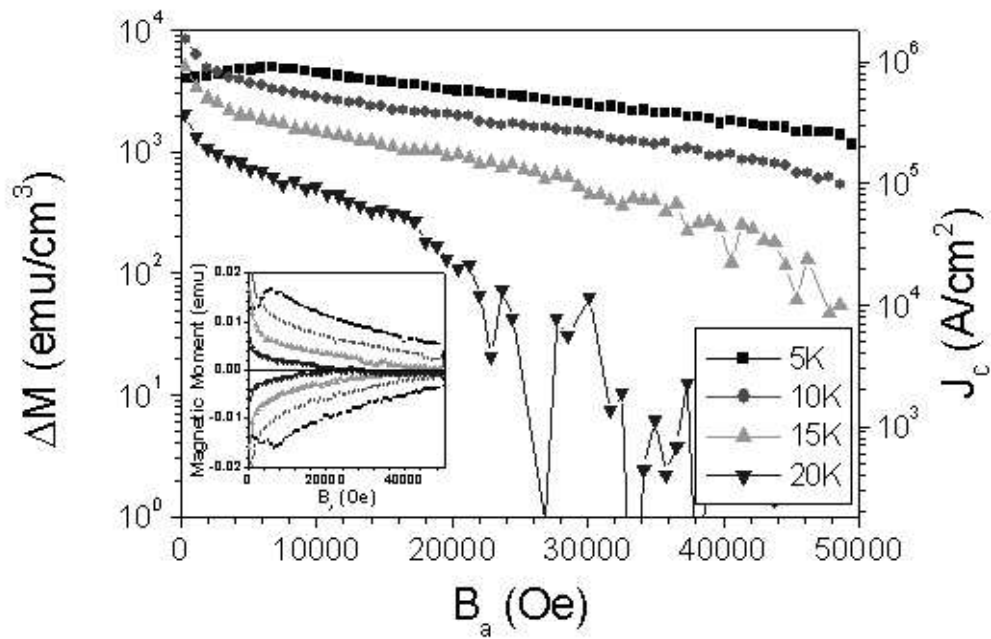

Fig. 5. Magnetization difference (ΔM) of film c as a function of applied field ($B_a$) at 5K, 10K, 15K, and 20K respectively. The right axis shows the calculated $J_c$ value from ΔM. The applied equation is shown in text. The inset figure shows magnetic hysteretic curves at the four temperatures.

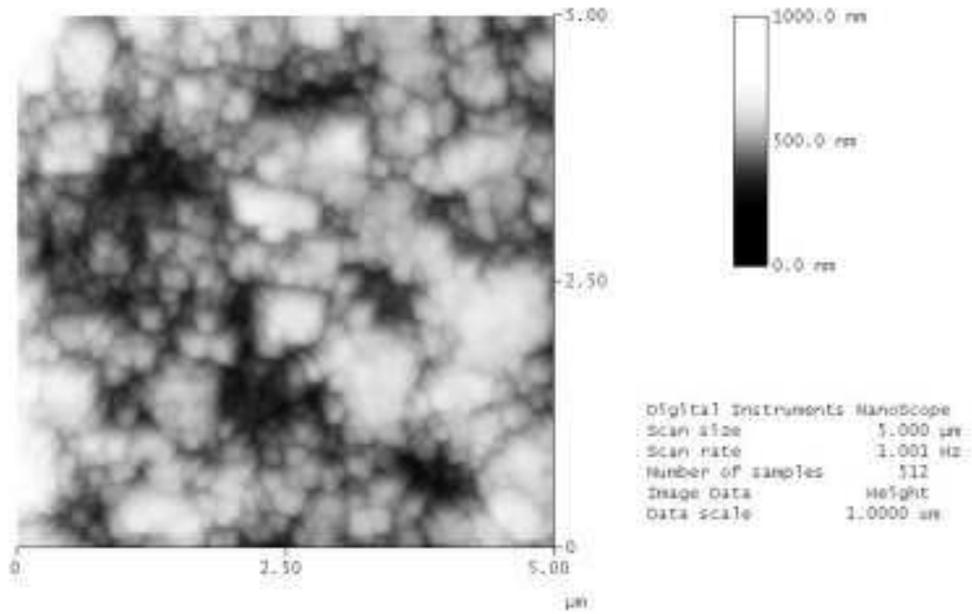

(a)

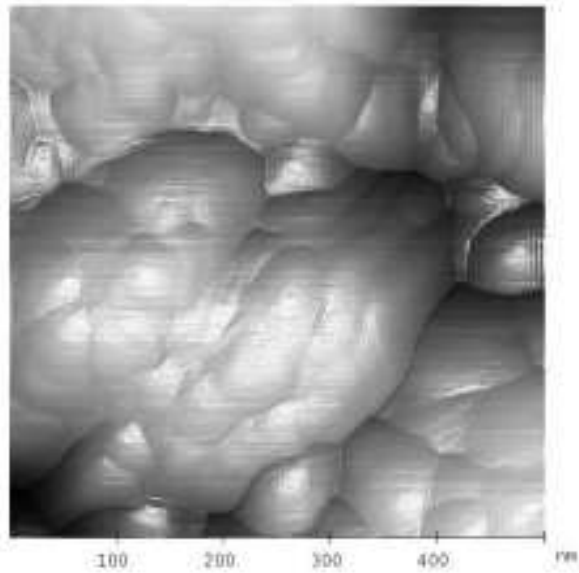

(b)

Fig. 6. AFM images of the MgB$_2$ films. (a) scale=5x5μm$^2$, Height image; (b) scale=500x500nm$^2$, Mixed height-illumination image.